\documentclass[a4paper,11pt]{article}
\usepackage{pos}

\title{Axion search with BabyIAXO in view of IAXO}

\newcommand{\UHEI}{1}
\newcommand{\UNIZAR}{2}
\newcommand{\CERN}{3}
\newcommand{\CEAIrfu}{4}
\newcommand{\INAF}{5}
\newcommand{\IJCLAB}{6}
\newcommand{\MIPT}{7}
\newcommand{\INR}{8}
\newcommand{\DTU}{9}
\newcommand{\ICCUB}{10}
\newcommand{\UBFQA}{11}
\newcommand{\PNPI}{12}
\newcommand{\UBonn}{13}
\newcommand{\USiegen}{14}
\newcommand{\BARRY}{15}
\newcommand{\JENA}{16}
\newcommand{\Columbia}{17}
\newcommand{\CEFCA}{18}
\newcommand{\SOLEIL}{19}
\newcommand{\RBI}{20}
\newcommand{\DESY}{21}
\newcommand{\CEAList}{22}
\newcommand{\ICREA}{23}
\newcommand{\MIT}{24}
\newcommand{\SLAC}{25}
\newcommand{\LLNL}{26}
\newcommand{\Mainz}{27}
\newcommand{\UCT}{28}

\newcommand{\UNIZARname}{Center for Astroparticles and High Energy Physics (CAPA), Universidad de Zaragoza, 50009 Zaragoza, Spain}
\newcommand{\BARRYname}{Physical Sciences, Barry University, 11300 NE 2nd Ave., Miami Shores, FL 33161, USA}
\newcommand{\MITname}{Massachusetts Institute of Technology, 77 Massachusetts Ave., Cambridge, MA 02139, USA}
\newcommand{\CEFCAname}{Centro de Estudios de F\'{\i}sica del Cosmos de Arag\'on, Plaza San Juan, Teruel, Spain}
\newcommand{\ICCUBname}{Institut de Ci\`encies del Cosmos, Universitat de Barcelona, Barcelona, Spain}
\newcommand{\UBFQAname}{Departament de F\'isica Qu\`antica i Astrof\'isica, Universitat de Barcelona, Barcelona, Spain}
\newcommand{\ICREAname}{ICREA, Barcelona, Spain}
\newcommand{\Mainzname}{Johannes Gutenberg University, Mainz, Germany}
\newcommand{\PNPIname}{Petersburg Nuclear Physics Institute - NRC Kurchatov Institute, Gatchina, 188300, Russia}
\newcommand{\RBIname}{Rudjer Bo\v{s}kovi\'{c} Institute, Bijeni\v{c}ka cesta 54, 10000 Zagreb, Croatia}
\newcommand{\UBonnname}{Physikalisches Institut der Universit\"at Bonn, Nussallee 12, 53115 Bonn, Germany}
\newcommand{\UCTname}{High Energy Physics, Cosmology \& Astrophysics Theory (HEPCAT) group, University of Cape Town, Private Bag, 7700 Rondebosch, South Africa}
\newcommand{\DESYname}{Deutsches Elektronen-Synchrotron DESY, Hamburg, Germany}
\newcommand{\CERNname}{CERN - European Organization for Nuclear Research, Geneva, Switzerland}
\newcommand{\UHEIname}{Heidelberg University, Kirchhoff Institute for Physics}
\newcommand{\LLNLname}{Lawrence Livermore National Laboratory, Livermore, CA, U.S.A.}
\newcommand{\Columbianame}{Columbia Astrophysics Laboratory, New York, U.S.A.}
\newcommand{\DTUname}{Technical University of Denmark, DTU Space Kgs. Lyngby, Denmark}
\newcommand{\SLACname}{SLAC National Accelerator Laboratory, Menlo Park, CA, U.S.A.}
\newcommand{\JENAname}{Institute  for  Optics  and  Quantum  Electronics,  Friedrich  Schiller  University  Jena,  Jena,  Germany}
\newcommand{\INAFname}{INAF, Italian National Institute for Astrophysics, Osservatorio Astronomico di Brera, Milano/Merate, Italy}
\newcommand{\INRname}{Institute  for Nuclear Research of the Russian Academy of Sciences, 60th October Anniversary Prospect 7A, 117312, Moscow, Russia}
\newcommand{\USiegenname}{Center for Particle Physics Siegen, Siegen University, Siegen, Germany}
\newcommand{\CEAIrfuname}{IRFU, CEA, Universit\'e Paris-Saclay, F-91191 Gif-sur-Yvette, France}
\newcommand{\CEAListname}{CEA, LIST, Laboratoire National Henri Becquerel (LNE-LNHB), F-91191 Gif-sur-Yvette, France}
\newcommand{\MIPTname}{Moscow Institute of Physics and Technology, Moscow, Russia}
\newcommand{\IJCLABname}{Universit\'{e} Paris-Saclay, CNRS/IN2P3, IJCLab, 91405 Orsay, France}
\newcommand{\SOLEILname}{Synchrotron SOLEIL, 91192 Gif-sur-Yvette, France}

\author[\UHEI]{A.~Abeln}
\author[\UNIZAR]{K.~Altenm\"uller}
\author[\CERN]{S.~Arguedas Cuendis}
\author[\CEAIrfu]{E.~Armengaud}
\author[\CEAIrfu]{D.~Atti\'e}
\author[\CEAIrfu]{S.~Aune}
\author[\INAF]{S.~Basso}
\author[\IJCLAB]{L.~Berg\'e}
\author[\CEAIrfu]{B.~Biasuzzi}
\author[\CERN]{P.~T.~C.~Borges~De~Sousa}
% ...

\emailAdd{javier.galan@unizar.es}

\abstract{Axions are a natural consequence of the Peccei-Quinn mechanism, the most compelling solution to the strong-CP problem. Similar axion-like particles (ALPs) also appear in a number of possible extensions of the Standard Model, notably in string theories. Both axions and ALPs are very well motivated candidates for Dark Matter, and in addition, they would be copiously produced at the sun’s core. A relevant effort during the last decade has been the CAST experiment at CERN, the most sensitive axion helioscope to-date. The International Axion Observatory (IAXO) is a large-scale 4th generation helioscope. As its primary physics goal, IAXO will look for solar axions or ALPs with a signal to background ratio of about 5 orders of magnitude higher than CAST. 
Recently the IAXO collaboration has proposed and intermediate experimental stage, BabyIAXO, conceived to test all IAXO subsystems (magnet, optics, detectors and sun-tracking systems) at a relevant scale for the final system and thus serve as pathfinder for IAXO but at the same time as a fully-fledged helioscope with record and relevant physics reach in itself with potential for discovery.
BabyIAXO was endorsed by the Physics Review committee of DESY last May 2019. Here we will review the status and prospects of BabyIAXO and its potential to probe the most physics motivated regions of the axion \& ALPs parameter space.}

\FullConference{
  40th International Conference on High Energy physics - ICHEP2020\\
  July 28 - August 6, 2020\\
  Prague, Czech Republic (virtual meeting)
}

\begin{document}
\maketitle

\section{Introduction}
The axion is a hypothetical particle that emerged as a natural solution to the \emph{strong CP problem}. The problem arises from the fact that the description of strong interactions given by the quantum chromodynamics (QCD) theory does not forbid such violation. Furthermore, since electroweak interactions violate CP it is difficult to understand why strong interactions do not. The Peccei-Quinn mechanism to restore CP conservation at the strong sector~\cite{Peccei:1977hh,Peccei:1977ur} developed into an interesting outcome, the axion, a new boson arising naturally in the theory~\cite{Weinberg:1977ma,Wilczek:1977pj}. The phenomenological properties of the axion are described through a unique parameter, the scale factor $f_a$, which must be much greater than the electroweak scale in order to circumvent the constrains imposed by accelerator based searches. Both, the couplings of axions to ordinary matter, and the axion mass, $m_a$, are inversely proportional to $f_a$. The parameter space of axions is further constrained by cosmological and astrophysical arguments, such as the evolution of stars or the dark matter content of the universe. As a consequence of its very weak couplings the axion results to be a long-lived particle, and therefore, under certain conditions it is considered to be a potential \emph{dark matter} candidate.

Two main theoretical models are considered as a reference for axion searches, \emph{hadronic axions} or \emph{KSVZ} axions~\cite{KimModified,Shifman:1979if}, and \emph{DSFZ} or \emph{GUT} axions~\cite{Dine:1981rt,Zhitnitsky:1980tq}, which do not couple to hadrons at tree level. Furthermore, a plethora of axion-like particles (ALPs) emerge in different extensions to the SM. Their similarity with standard QCD axions makes that IAXO will be able to exploit its physics program to explore ALPs hinted regions other than the theoretical QCD models~\cite{Irastorza:2018dyq,Graham:2015ouw}

\section{BabyIAXO and IAXO Physics potential}

\begin{figure}[th]
\begin{center}
\includegraphics[height=7.5cm]{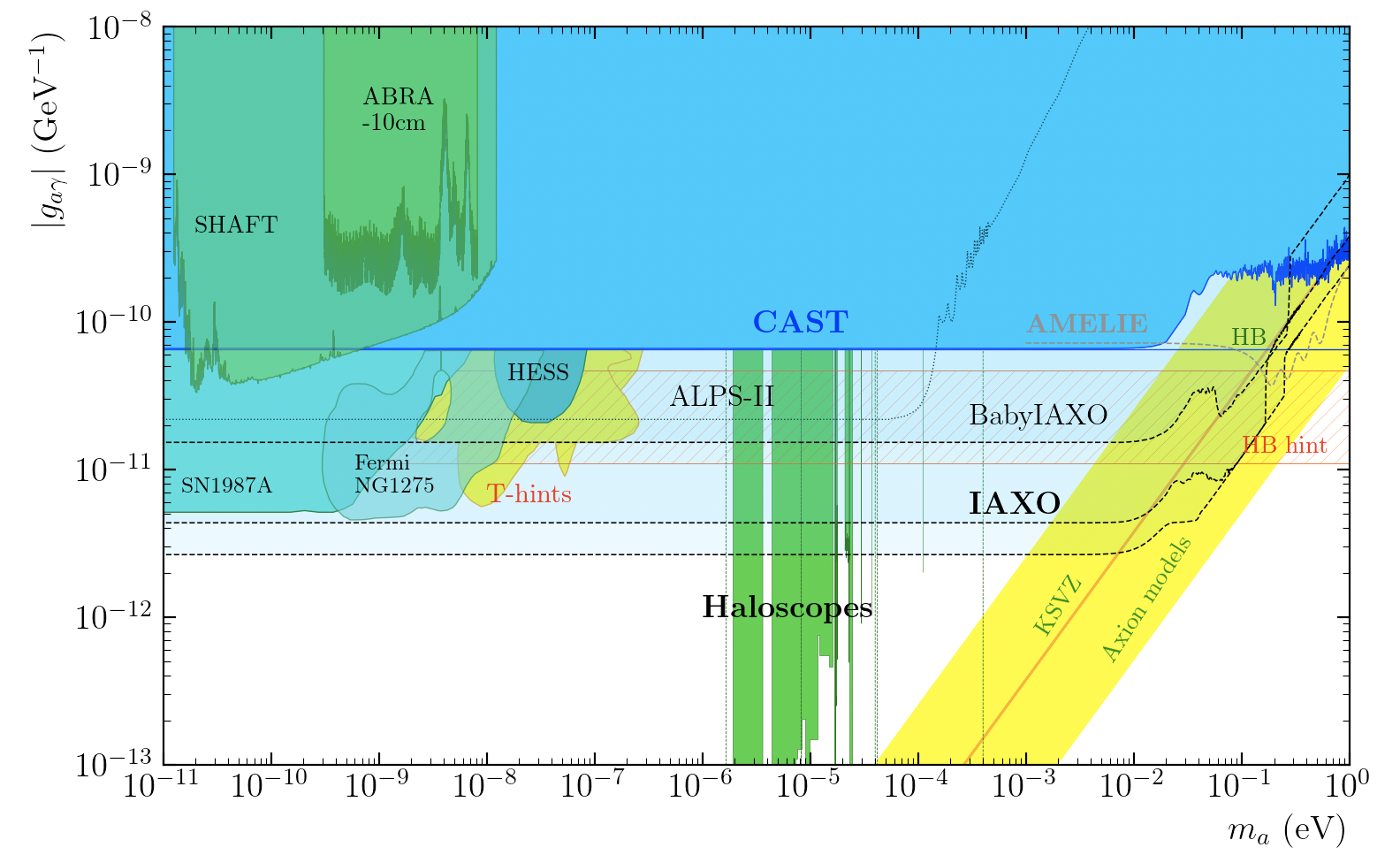}
\caption{\label{fig:exclusion} The prospects of the different IAXO setups and its sensitivity to $g_{a\gamma}$ as a function of the axion mass, $m_a$, including different hints and models described in the text~\cite{Irastorza:2018dyq}.}
\end{center}
\end{figure}

IAXO is a 4th generation helioscope that will push the sensitivity to solar axions to unprecedented levels, surpassing those of the most sensitive axion helioscope up to date, the CERN Axion Solar Telescope (CAST)~\cite{Anastassopoulos:2017ftl}. The final IAXO configuration will allow to explore a vast region of the axion parameter space, covering theoretically QCD favored regions, stellar cooling hints as the horizontal branch stars~\cite{Straniero_modified} or white dwarf (WD) cooling anomalies~\cite{Melendez:2012iq}, and the universe gamma-rays transparency anomaly reported by HESS~\cite{Aharonian:2005gh} and MAGIC~\cite{Aliu:2008ay}, between other hints. Figure~\ref{fig:exclusion} shows those hints and the sensitivities reached by BabyIAXO and IAXO.  BabyIAXO, being a prototype of the full IAXO, will cover itself an interesting new unexplored region in the parameter space, reaching $g_{a\gamma} \sim 1.5\times 10^{-11}$\,GeV$^{-1}$ for masses up to $0.25$\,eV, including the favored QCD axion region. See reference~\cite{Armengaud_2019} for a full review.

\section{Design and construction of BabyIAXO}

BabyIAXO will be hosted by the DESY research facility after the recent endorsement by the DESY Physics Review committe in May 2019. Most of the BabyIAXO systems (cryogenics, magnet, optics, detector systems) are in an advanced mature state and moving towards the construction phase. BabyIAXO will benefit from the existence of a Medium-Sized Telescope (MST)~\cite{Garczarczyk:2015zya} instrument, commissioned  by the Chrerenkov Telescope Array (CTA) group at DESY, to serve as a dedicated driving system for BabyIAXO. The most critical element of the system is the magnet system which is being designed and commissioned at CERN~\cite{Abeln:2020ywv}.

 The figure of merit (FOM) of the experiment can be factorized on different experimental systems~\cite{Armengaud_2019}, such as the magnet, $f_{M}=B^2L^2A$, the combination of optic and detectors systems, $f_{DO}=\frac{\epsilon_d\epsilon_o}{\sqrt(ba)}$, and tracking exposure, $f_T=\sqrt(\epsilon_tt)$. The total FOM, $f_{FOM}=f_Mf_{DO}f_T$, will be roughly proportional to the expected number of photons converted from axions that can be detected, and that will be 5 orders of magnitude higher than present measurements for the IAXO+ prospects, being IAXO+ the ultimate reach of the IAXO R\&D program where technology is pushed to the limits. Table~\ref{tab:parameters} shows a summary of the experimental setup parameters that apply to each sensitivity plot shown previously in~Figure~\ref{fig:exclusion}.


\begin{table}[th]
\centering
\begin{tabular}{cccc}
    \bf{Parameter [units]} &  \bf{BabyIAXO} & \bf{IAXO}  &   \bf{IAXO+}   \\ \hline 
   \emph{B [T]}   &    $\sim$2        &  $\sim$2.5    &  $\sim$3.5     \\
   \emph{L [m]}   &    10        &  22    &  22     \\ 
   \emph{A [m$^2$]}   &    0.77        &  2.3    &  3.9     \\ \hline
   \emph{f$_M$ [T$^2$m$^4$]}   &    $\sim$230        &  $\sim$6000    &  $\sim$64000     \\ \hline 
   
\rule{0pt}{4ex}  
   
   \emph{b [cm$^{-2}$\,keV$^{-1}$s$^{-1}$]}   &    10$^{-7}$       &  10$^{-8}$    &  10$^{-9}$     \\
   \emph{$\epsilon_d$}   &    0.7        &  0.8    &  0.8     \\ 
   \emph{$\epsilon_o$}   &    0.35        &  0.7    &  0.7    \\ 
   \emph{a [cm$^{2}$]}   &    2$\times$0.3       &  8$\times$0.15    &  8$\times$0.15     \\ \hline 
   \emph{f$_{DO}$ [keV$^{1/2}$s$^{1/2}$]}   &    1000        &  5112    &   16166    \\ \hline 
 
\rule{0pt}{4ex}  

   \emph{$\epsilon_t$}   &   0.5      &  0.5    &  0.5     \\ 
   \emph{t [year]}   &    1.5        &  3    &  5     \\\hline 
   \emph{f$_T$ [year$^{1/2}$]}   &    0.87        &  1.22    &  1.58     \\ \hline 

\end{tabular}
\caption{\label{tab:parameters} A comparison of the relevant experimental parameters from BabyIAXO to IAXO+.}
\end{table}

\begin{figure}[ht]
\begin{center}
\includegraphics[height=6cm]{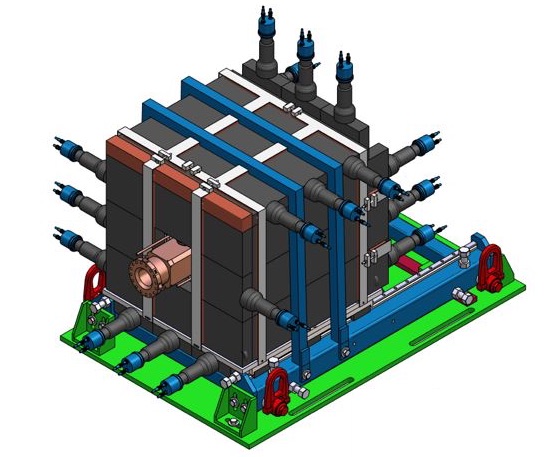}
\caption{\label{fig:shielding} The undergoing CAD design for the shielding of detectors including an active VETO system .}
\end{center}
\end{figure}

An important upgrade for IAXO is the magnet contribution to the $f_{FOM}$, that provides to the magnetic field configuration with a much bigger aperture, and additional bores, with respect to its predecessor, CAST. After the magnetic field, and driving system (which provides also a longer tracking exposure, $\sim$18\,hours-per-day) have been fixed, the next system to enhance axion searches is the x-ray detector system which is under active design, as shown in Figure~\ref{fig:shielding}.

\begin{figure}[hb]
\begin{center}
\includegraphics[height=3.8cm]{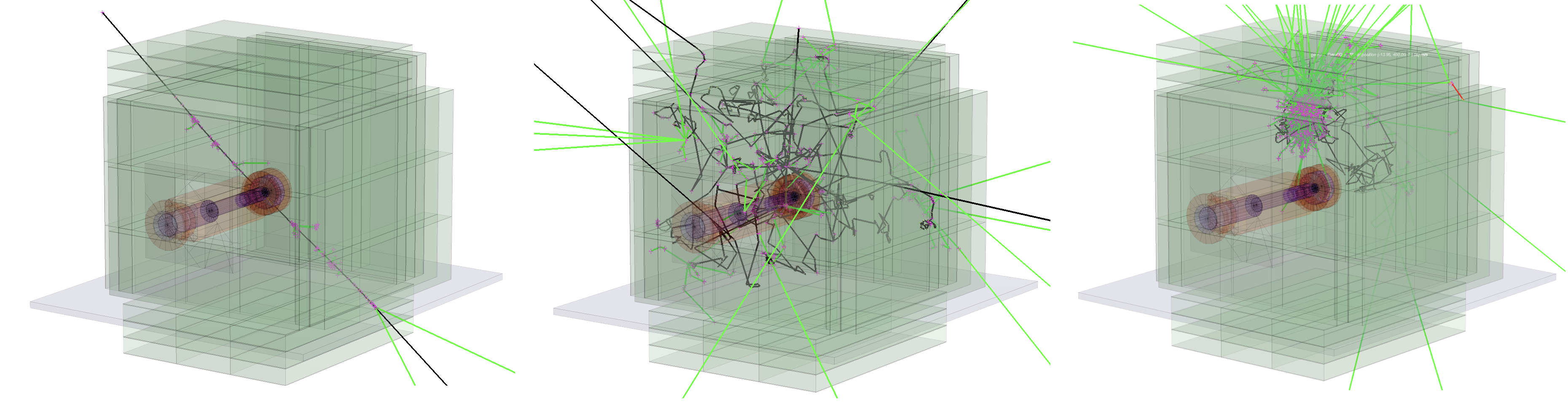}
\caption{\label{fig:Geant4MC} \emph{Geant4} Monte Carlo simulations of the detector, shielding VETO systems using RESTSoft~\cite{REST_modified}. From left to right, a simulated muon, neutron and high-energy gamma. Green lines describe the trajectories of gammas and black lines of neutrons.}
\end{center}
\end{figure}

In that sense, extensive studies are taking place in order to better understand the nature of the background to achieve the low levels required. The optimization of the active VETO tagging system to different sources, as cosmic neutrons, gammas and muons (see Figure~\ref{fig:Geant4MC}) is crucial for the success of IAXO, in order to achieve the background level of 10$^{-8}$keV$^{-1}$cm$^{-2}$s$^{-1}$, and ultimately a level of 10$^{-9}$ leading to the detection of just $\sim$0.5 background counts if the prospects for IAXO+ scenario are achieved. BabyIAXO will be no doubt an excellent center of operations to demonstrate and optimize all the final systems that will be finally used to construct IAXO.

\bibliographystyle{JHEP} 
\bibliography{bib/IAXObib,bib/detectors,bib/IAXObib2}
%

\end{document}